\documentclass[aps,prl,preprint,showpacs,groupedaddress]{revtex4}
\usepackage {amssymb}
\usepackage {amsmath}
\usepackage {graphicx}
\usepackage {longtable}

\begin{document}
\title{Energy pseudogaps and structural transitions in metals}

\author{Fedor V.Prigara}
\affiliation{Institute of Physics and Technology,
Russian Academy of Sciences,\\
21 Universitetskaya, Yaroslavl 150007, Russia}
\email{fprigara@recnti.uniyar.ac.ru}

\date{\today}

\begin{abstract}

It is shown that a structural transition in a metal is associated with the
opening of the energy pseudogap in the electronic spectrum in a
low-temperature phase below the transition temperature. A relation between
the magnitude of the energy pseudogap in the d-band of transition metals (at
zero temperature) and the structural transition temperature is similar to a
relation between the magnitude of the superconducting gap and the
superconducting transition temperature in low-temperature and
high-temperature superconductors. A relation between the magnitude of the
energy pseudogap and the structural transition temperature in sp-metals is
similar to a relation between the bandgap width (at zero temperature) and
the metal-insulator transition temperature in semiconductors.

\end{abstract}


\maketitle

The superconducting transition is associated with the opening of
the energy gap in the electronic spectrum at the Fermi level [1].
The magnitude of the superconducting gap $2\Delta _{sc} \left( {0}
\right)$ at zero temperature ($T = 0K$) is proportional to the
superconducting transition temperature $T_{c} $, the
proportionality coefficient for high-temperature superconductors
being twice with respect to those for low-temperature
superconductors [2]. Here we show that a similar relation with the
same coefficients exists between the magnitude $\Delta \left( {0}
\right)$ of the energy pseudogap in the d-band of transition
metals and the structural transition temperature $T_{s} $. For
sp-metals, a relation between the magnitude of the energy
pseudogap (corresponding to a relative shift of the s- and p-bands
with respect to each other) and the structural transition
temperature is different and similar to a relation between the
bangap width at zero temperature and the metal-insulator
transition temperature in semiconductors. The bandgap width in the
undoped insulating phase of high-temperature superconductors is
close to the magnitude of the energy pseudogap in metals which
form charge-ordered layers in these superconductors [3,4].

The magnitude $2\Delta _{sc} \left( {0} \right)$ of the superconducting gap
at zero temperature ($T = 0K$) is related to the superconducting transition
temperature $T_{c} $ by the equation [2]

\begin{equation}
\label{eq1}
2\Delta _{sc} \left( {0} \right) = \alpha _{P} \alpha k_{B} T_{c} ,
\end{equation}

\noindent
where $\alpha _{P} $ is the atomic relaxation constant, $\alpha _{P} = 3/16$
for low-temperature superconductors and $\alpha _{P} = 3/8$ for
high-temperature superconductors, $\alpha = 18$ is a constant, and $k_{B} $
is the Boltzmann constant.

The difference of the free energies of the normal state and the
superconducting state, respectively, at zero temperature is given by the
formula

\begin{equation}
\label{eq2}
F_{n} - F_{s} = \frac{{1}}{{2}}N\left( {E_{F}}  \right)\Delta _{sc}^{2}
\left( {0} \right) - E_{fe} ,
\end{equation}

\noindent where $N\left( {E_{F}}  \right)$ is the density of
states at the Fermi level, and $E_{fe} $ is an increase in the
energy of electrons caused by a general contraction of the
conduction band produced by a ferroelastic distortion associated
with the superconducting transition (the ferroelastic energy) [3].
The first term on the right side of the equation (\ref{eq2}) is
the condensation energy $E_{0} $ [1].

The density of states at the Fermi level $N\left( {E_{F}}  \right)$ can be
determined from the electronic specific heat coefficient $\gamma $ [5],

\begin{equation}
\label{eq3}
\gamma = \frac{{2\pi ^{2}}}{{3}}N\left( {E_{F}}  \right)k_{B}^{2} ,
\end{equation}

\noindent
so that the equation (\ref{eq2}) takes a form

\begin{equation}
\label{eq4}
F_{n} - F_{s} = \frac{{3}}{{16\pi ^{2}}}\left( {\alpha _{P} \alpha}
\right)^{2}\gamma T_{c}^{2} - E_{fe} .
\end{equation}

For type I superconductors, the difference of the free energies of the
normal state and the superconducting state is related to the critical field
$H_{c} \left( {0} \right)$ at zero temperature by a thermodynamic formula
[1]

\begin{equation}
\label{eq5}
F_{n} - F_{s} = \frac{{H_{c}^{2} \left( {0} \right)}}{{8\pi} }V_{m} ,
\end{equation}

\noindent
where $V_{m} $ is the molar volume.

A comparison of the equations (\ref{eq4}) and (\ref{eq5}) with experimental data for the
electronic specific heat coefficient $\gamma $ and the critical field $H_{c}
\left( {0} \right)$ at zero temperature for type I superconductors shows
that the ferroelastic energy $E_{fe} $ is normally small in low-temperature
superconductors with a sufficiently high $T_{c} $. For example, the $E_{fe}
/E_{0} $ ratio is about 0.06 for Ru and Cd, so that the equation (\ref{eq4}) gives
for low-temperature superconductors with the atomic relaxation constant
$\alpha _{P} = 3/16$

\begin{equation}
\label{eq6}
F_{n} - F_{s} \approx 0.216\gamma T_{c}^{2} .
\end{equation}

If the ferroelastic energy $E_{fe} $ exceeds the condensation energy $E_{0}
$, then the superconducting transition does not occur [3].

The magnitude of the superconducting gap can be determined by means of
scanning tunneling spectroscopy. The tunneling spectra of Fe(Se,Te) [6] show
that there are two superconducting phases in this compound. The surface
phase corresponds to a low-temperature superconductivity with the
superconducting transition temperature $T_{c1} = 11K$, the superconducting
gap $2\Delta _{1} \left( {0} \right) = 3.4meV$, and the atomic relaxation
constant $\alpha _{P} = 3/16$. The bulk phase corresponds to a
high-temperature superconductivity with the superconducting transition
temperature $T_{c2} = 14.5K$, the superconducting gap $2\Delta _{2} \left(
{0} \right) = 8.4meV$, and the atomic relaxation constant $\alpha _{P} =
3/8$. The surface phase has presumably a tetragonal crystal structure, and
the bulk phase has a monoclinic structure [7]. The lattice constant of the
surface phase $a = 0.38nm$ corresponds to the tetragonal phase.

The d-band in transition metals is normally split into subbands
due to structural transitions or ferroelastic distortions
associated with ferromagnetic or antiferromagnetic ordering [3].
If the full magnitude $\Delta \left( {0} \right)$ of the energy
pseudogap is related to the structural transition temperature
$T_{s} $ by the equation

\begin{equation}
\label{eq7} \Delta \left( {0} \right) = \alpha _{P} \alpha k_{B}
T_{s} ,
\end{equation}

\noindent
then the difference $E_{2} - E_{1} $ of the energies of a high-temperature
phase and a low-temperature phase, respectively, can be estimated from the
formula similar to the equation (\ref{eq4}),

\begin{equation}
\label{eq8}
E_{2} - E_{1} \cong \frac{{3}}{{16\pi ^{2}}}\left( {\alpha _{P} \alpha}
\right)^{2}\gamma T_{s}^{2} .
\end{equation}

The difference of the energies of a high-temperature phase and a
low-temperature phase at the structural transition temperature $T_{s} $ is
equal to the latent heat of the transition, $\Delta H$,

\begin{equation}
\label{eq9}
E_{2} - E_{1} = \Delta H.
\end{equation}

A comparison of the equations (\ref{eq8}) and (\ref{eq9}) with experimental data for
electronic specific heat coefficient $\gamma $ and the latent heat of the
structural transition $\Delta H$ in transition metals [8] shows that the
atomic relaxation constant is $\alpha _{P} = 3/8$ for the hcp-bcc transition
in Ti, Zr, and Hf, and also for the fcc-bcc transition in Ca and for the
hcp-bcc transition in Sr. For most of structural transformations in
transition metals, the atomic relaxation constant is $\alpha _{P} =
3/16$,for example, in the case of structural transitions in Mn, Fe, Co, and
in rare earth metals.

The fcc-bcc transition in Ca and the hcp-bcc transition in Sr are
d-band type transitions. A partial filling of the d-band in
alkaline earth metals is supported by an enhanced paramagnetic
susceptibility exceeding the Pauli paramagnetic susceptibility,
contrary to the case of alkali metals.

In the case of the hcp-bcc transition in Ti at $T = 1166K$, the electronic
specific heat coefficient is $\gamma = 3.45mJmol^{ - 1}K^{ - 2}$, and the
equation (\ref{eq8}) with the atomic relaxation constant $\alpha _{P} = 3/8$ gives
$E_{2} - E_{1} \cong 4.2kJmol^{ - 1}$, whereas the latent heat of the
transition is $\Delta H = 4.0kJmol^{ - 1}$. In the case of the $\alpha -
\beta $ transition in Mn at $T_{s} = 973K$, the electronic specific heat
coefficient is $\gamma = 10.6mJmol^{ - 1}K^{ - 2}$,and the equation (\ref{eq8}) with
the atomic relaxation constant $\alpha _{P} = 3/16$ gives $E_{2} - E_{1}
\cong 2.18kJmol^{ - 1}$, whereas the latent heat of the transition is
$\Delta H = 2.23kJmol^{ - 1}$. The equation (\ref{eq7}) gives the magnitude of the
energy pseudogap $\Delta = 0.68eV$ for the hcp-bcc transition in Ti, and
$\Delta _{1} = 0.28eV$ for the $\alpha - \beta $ transition in Mn.

The hcp-bcc transition in Be is an sp-band type transition . In this case,
the full magnitude $\Delta \left( {0} \right)$ of the energy pseudogap is
determined by the equation

\begin{equation}
\label{eq10}
\Delta \left( {0} \right) = \alpha k_{B} T_{s} ,
\end{equation}

\noindent
which is similar to a relation between the bandgap width $E_{g} \left( {0}
\right)$ at zero temperature and the metal-insulator transition temperature
$T_{MI} $ in semiconductors [3],

\begin{equation}
\label{eq11}
E_{g} \left( {0} \right) = \alpha k_{B} T_{MI} .
\end{equation}

The difference $E_{2} - E_{1} $ of the energies of a high-temperature phase
and a low-temperature phase, respectively, for sp-band transitions is given
by the equation

\begin{equation}
\label{eq12}
E_{2} - E_{1} \cong \frac{{3}}{{16\pi ^{2}}}\alpha ^{2}\gamma T_{s}^{2}
\approx 6.1\gamma T_{s}^{2} .
\end{equation}

In the case of the hcp-bcc transition in Be at $T_{s} = 1450K$, the
electronic specific heat coefficient is $\gamma = 0.22mJmol^{ - 1}K^{ - 2}$,
and the equation (\ref{eq12}) gives $E_{2} - E_{1} \cong 2.8kJmol^{ - 1}$, whereas
the latent heat of the transition is $\Delta H = 2.1kJmol^{ - 1}$. The
magnitude of the energy pseudogap in Be is $\Delta = 2.25eV$, according to
the equation (\ref{eq10}). The sp-band transition in Be produces a low density of
states at the Fermi level in the low-temperature hcp phase and, as a
consequence, its diamagnetism.

The fcc-hcp transition in Sr is also an sp-band transition, however, the
magnitude of the energy pseudogap in Sr is much smaller than in Be, $\Delta
= 0.8eV$. The fcc-hcp transition in Ce at $T_{s} = 115K$ is an sp-band
transition. Other structural transitions in Ce, the hcp-fcc transition and
the fcc-bcc transition, belong to a d-band type with the atomic relaxation
constant $\alpha _{P} = 3/16$.

Structural transitions of a d-band type in d-metals and f-metals
invoke also a change in the structure of the sp-bands, so that the
location of the Fermi level is different in a high-temperature
phase and a low-temperature phase. The $\beta $ phase of Mn (of
the Hume-Rothery type) is `nonmagnetic' (the Mn atoms have no
magnetic moments) [9], which corresponds to a fully filled 6-state
d-subband and an empty 4-state d-subband. In the $\alpha $ phase
of Mn (of the Frank-Kasper type), the 6-state d-subband is split.
$\alpha - Mn$ exhibits antiferromagnetic ordering with the Neel
temperature $T_{N} = 95K$.

The magnitude $\Delta _{AFM} $ of the antiferromagnetic pseudogap is
determined by the formula similar to the equation (\ref{eq10}) [2],

\begin{equation}
\label{eq13}
\Delta _{AFM} = \alpha k_{B} T_{N} .
\end{equation}

This equation gives for the $\alpha $ phase of Mn $\Delta _{AFM} = 0.148eV$,
so that there is a relation

\begin{equation}
\label{eq14}
2\Delta _{AFM} \cong \Delta _{1} ,
\end{equation}

\noindent
where $\Delta _{1} = 0.284eV$ is the energy pseudogap associated with the
$\alpha - \beta $ transition in Mn.

For antiferromagnetic fluctuations in the superconducting phase, there is a
similar relation between the magnitude $\Delta _{AFM}^{\ast}  $ of the
antiferromagnetic pseudogap and the magnitude $2\Delta _{sc} \left( {0}
\right)$ of the superconducting gap [2],

\begin{equation}
\label{eq15}
\Delta _{AFM}^{ *}  \cong 2\Delta _{sc} \left( {0} \right).
\end{equation}

In view of the equations (\ref{eq13}) and (\ref{eq1}), this relation gives the temperature
$T_{AFM}^{ *}  $ of the maximum of antiferromagnetic fluctuations in the
superconducting phase in the form [2]

\begin{equation}
\label{eq16}
T_{AFM}^{ *}  \approx \alpha _{P} T_{c} .
\end{equation}

At the temperature $T_{AFM}^{ *}  $, there is a peak (a shoulder feature) in
the temperature dependence of the specific heat, for example, in the
low-temperature superconductor $KFe_{2} As_{2} $ [10] with the
superconducting transition temperature $T_{c} = 3.7K$ and the atomic
relaxation constant $\alpha _{P} = 3/16$. A similar peak in the specific
heat at the temperature $T_{AFM}^{ *}  $ was observed in the
high-temperature iron pnictide superconductor $Ba_{1 - x} K_{x} Fe_{2}
As_{2} $ [11] with the superconducting transition temperature $T_{c} = 38K$
and the atomic relaxation constant $\alpha _{P} = 3/8$.

The bandgap width $E_{g} \left( {0} \right)$ in the undoped insulating phase
of high-temperature superconductors is normally close to the magnitude of
the energy pseudogap in metals which form charge-ordered layers in these
superconductors. The bandgap width $E_{g} \left( {0} \right)$ is determined
by the equation (\ref{eq11}). In $YBa_{2} Cu_{3} O_{y} $ with the superconducting
transition temperature $T_{c} = 92K$, the metal-insulator transition
temperature in the undoped phase is $T_{MI} = 410K$ [12], so that $E_{g}
\left( {0} \right) = 0.635eV$. The magnitude $\Delta $ of the energy
pseudogap associated with the hcp-fcc transition in Y at $T_{s} = 1753K$,
according to the equation (\ref{eq7}) with the atomic relaxation constant $\alpha
_{P} = 3/16$, is $\Delta = 0.51eV$.

In $La_{2} CuO_{4 + \delta}  $ with the superconducting transition
temperature $T_{c} = 45K$ [13], the metal-insulator transition temperature
in the undoped phase is $T_{MI} = 240K$, and $E_{g} \left( {0} \right) =
0.37eV$. The magnitude $\Delta $ of the energy pseudogap associated with the
fcc-bcc transition in La at $T_{s} = 1133K$ is $\Delta = 0.33eV$ (the atomic
relaxation constant is $\alpha _{P} = 3/16$).

In the possible high-temperature superconductor $Ca_{2} RuO_{4 + \delta}  $,
the metal-insulator transition temperature in the undoped phase is $T_{MI} =
357K$ [14], so that $E_{g} \left( {0} \right) = 0.55eV$. The magnitude
$\Delta $ of the energy pseudogap associated with the fcc-bcc transition in
Ca at $T_{s} = 716K$ is $\Delta = 0.42eV$ (the atomic relaxation constant is
$\alpha _{P} = 3/8$).

In the electron-doped iron pnictide superconductor $Ba\left( {Fe_{1 - x}
Co_{x}}  \right)_{2} As_{2} $ with the superconducting transition
temperature $T_{c} = 23K$ [15], the metal-insulator transition temperature
in the undoped phase is $T_{MI} \cong 150K$ (an extrapolated value) [3, 16],
and $E_{g} \left( {0} \right) = 0.23eV$. The magnitude $\Delta $ of the
energy pseudogap associated with the hcp-fcc transition in Co at $T_{s} =
700K$ is $\Delta = 0.20eV$ (the atomic relaxation constant is $\alpha _{P} =
3/16$).

To summerize, we show that structural transitions in metals are associated
with changes in their electronic structure. We obtain a relation between the
magnitude of the energy pseudogap in a low-temperature phase and the
structural transition temperature for sp-band type and d-band type
transitions. In the case of d-band type transitions, this relation is
similar to a relation between the magnitude of the superconducting gap at
zero temperature and the superconducting transition temperature in
low-temperature and high-temperature superconductors. The sp-band type
transitions in metals are similar to the metal-insulator transition. The
energy pseudogaps in metals which form charge-ordered layers in
high-temperature superconductors determine the bandgap width in the undoped
insulating phase of these superconductors.

\begin{center}
---------------------------------------------------------------
\end{center}

[1] D.Saint-James, G.Sarma, and E.J.Thomas, \textit{Type II
Superconductivity} (Pergamon Press, Oxford, 1969).

[2] F.V.Prigara, arXiv:0708.1230 (2007).

[3] F.V.Prigara, arXiv:1011.2359 (2010).

[4] K.Kudo, Y.Nishikubo, and M.Nohara, J. Phys. Soc. Jpn. \textbf{79},
123710 (2010).

[5] J.S.Blakemore, \textit{Solid State Physics} (W.B.Saunders Company,
Philadelphia, 1969).

[6] T.Hanaguri, S.Niitaka, K.Kuroki, and H.Takagi, Science \textbf{328}, 474
(2010).

[7] A.K.Stemshorn, Y.K.Vohra, P.M.Wu, F.C.Hsu, Y.L.Huang, M.K.Wu, and
K.W.Yeh, arXiv:1105.0908 (2011).

[8] E.M.Sokolovskaya and L.S.Guzei, \textit{Metal Chemistry} (Moscow
University Press, Moscow, 1986).

[9] R.J.Weiss, \textit{Solid State Physics for Metallurgists} (Pergamon
Press, Oxford, 1963).

[10] J.S.Kim, E.G.Kim, G.R.Stewart, X.H.Chen, and X.F.Wang, Phys. Rev. B
\textbf{83}, 172502 (2011).

[11] J.K.Dong, L.Ding, H.Wang, X.F.Wang, T.Wu, X.H.Chen, and S.Y.Li, New J.
Phys. \textbf{10}, 123031 (2008).

[12] Y.Wang and N.P.Ong, Proc. Nat. Acad. Sci. USA \textbf{98}, 11091
(2001).

[13] T.Hirayama, M.Nakagawa, and Y.Oda, Solid State Commun. \textbf{113},
121 (1999).

[14] H.Fukazawa, Nakatsuji, and Y.maeno, Physica B \textbf{281-282}, 613
(2000).

[15] S.L.Bud'ko, N.Ni, and P.C.Canfield, Phys. Rev. B \textbf{79}, 220516(R)
(2009).

[16] J.-H. Chu, J.G.Analytis, K.De Greve, P.L.McMahon, Z.Islam, Y.Yamamoto,
and I.R.Fisher, Science \textbf{329}, 824 (2010).

\end{document}